\documentclass[12pt]{extarticle}
\usepackage[utf8]{inputenc}
\usepackage{graphicx}
\usepackage{float}

\floatstyle{plaintop}
\restylefloat{table}

\usepackage[colorlinks]{hyperref}
\hypersetup{linkcolor=blue,citecolor=blue,filecolor=blue,urlcolor=blue}

\usepackage{authblk}
\usepackage{xcolor}
\usepackage{amsmath}

\usepackage{nameref}

\usepackage[margin=1.2in]{geometry}
\usepackage[font={small}, margin=0.75cm, labelfont=bf]{caption}

\usepackage[numbers]{natbib}

\usepackage{url}

\title{Carbonate-Silicate Cycle Predictions of Earth-like Planetary Climates
and Testing the Habitable Zone Concept}

\author[1,2,3,*]{Owen R. Lehmer}
\author[2,3]{David C. Catling}
\author[3,4,5]{Joshua Krissansen-Totton}
\affil[1]{MS 239-4, Space Science Division, NASA Ames Research Center, Moffett Field, CA, 94035, USA}
\affil[2]{Dept. Earth and Space Sciences/Astrobiology Program, Box 351310, University of Washington, Seattle, WA, 98195, USA}
\affil[3]{Virtual Planetary Laboratory at the University of Washington, Seattle, WA, 98195}
\affil[4]{Dept. Astronomy and Astrophysics, MS UCO/Lick Observatory, 1156 High Street, Santa Cruz, CA, 95064, USA}
\affil[5]{NASA Sagan Fellow}
\affil[*]{owen.r.lehmer@nasa.gov}

\begin{document}

\maketitle

\begin{abstract}
In the conventional habitable zone (HZ) concept, a CO$_{2}$-H$_2$O greenhouse
maintains surface liquid water. Through the water-mediated carbonate-silicate
weathering cycle, atmospheric CO$_{2}$ partial pressure (pCO$_{2}$) responds to
changes in surface temperature, stabilizing the climate over geologic
timescales. We show that this weathering feedback ought to produce a log-linear
relationship between pCO$_{2}$ and incident flux on Earth-like planets in the
HZ. However, this trend has scatter because geophysical and physicochemical
parameters can vary, such as land area for weathering and CO$_2$ outgassing
fluxes. Using a coupled climate and carbonate-silicate weathering model, we
quantify the likely scatter in pCO$_2$ with orbital distance throughout the HZ.
From this dispersion, we predict a two-dimensional
relationship between incident flux and pCO$_2$ in the HZ and show that it
could be detected from at least 83 ($2{\sigma}$) Earth-like exoplanet
observations. If fewer Earth-like exoplanets are observed, testing the
HZ hypothesis from this relationship could be difficult.

\end{abstract}

\section{Introduction} \label{sec:intro}
Newton first alluded to the concept of a stellar habitable zone (HZ) in his
1687 Principia\cite{newton_philosophiae_1687} by noting that Earth's
liquid water would vaporize or freeze at the orbits of Mercury and Saturn,
respectively\cite{cohen_principia_1999}. Later, Whewell noted that ``the
Earth's orbit is in the temperate zone of the Solar
System''\cite{whewell_plurality_1853}. Since then, the definition of the
stellar HZ has been refined, reaching its modern incarnation based on climate
models \cite{kasting_earths_1993,catling_atmospheric_2017}. 

Current HZ calculations \cite{kopparapu_habitable_2013} find that around a
Sun-like star, an Earth-like planet could remain habitable between 0.97 and
1.70 AU. The inner edge of the HZ is set by loss of surface water and the outer
edge is set by the maximum greenhouse of a CO$_2$ atmosphere where
extensive CO$_2$ condensation and increased Rayleigh scattering prevent any
further greenhouse warming from CO$_2$
\cite{kasting_habitable_1993,kopparapu_habitable_2013}. This definition of the
HZ only considers H$_2$O and CO$_2$ as greenhouse gases, so Earth-like planets
warmed by other greenhouse gases (e.g. H$_2$ or CH$_4$) could remain habitable
at bigger orbital distances
\cite{stevenson_life-sustaining_1999,seager_exoplanet_2013,catling_atmospheric_2017}.
However, CH$_4$-rich atmospheres in the HZ may not be possible without life to
generate substantial CH$_4$
\cite{krissansen-totton_disequilibrium_2018,wogan_abundant_2020}. In addition, more
complex climate models have shown the HZ might extend to smaller orbital
distances, perhaps interior to Venus' orbit, with appropriate planetary
conditions
\cite{abe_habitable_2011,zsom_toward_2013,yang_strong_2014,way_was_2016}.

Residing within the HZ does not guarantee habitable surface conditions.
Crucially, greenhouse gas abundances (and planetary albedo) must conspire to
produce clement surface conditions. For example, by most estimates, Mars
resides within the Sun's HZ but is not habitable because there is insufficient
greenhouse warming from CO$_2$, in part because of the lack of volcanic
outgassing of CO$_2$. Thus, considering the planetary
processes that control atmospheric CO$_2$ abundances on Earth-like planets
in the HZ is necessary to constrain planetary habitability.

The prevailing hypothesis is that CO$_2$ levels are controlled by a weathering
thermostat \cite{walker_negative_1981}. This can explain why Earth has
maintained a clement surface throughout its history despite the ${\sim}$30\%
brightening of the Sun over the past ${\sim}$4.5 Gyr \cite{sagan_earth_1972,
tajika_evolution_1992, tajika_degassing_1993, sleep_carbon_2001,
krissansen-totton_constraining_2018}.  The changing luminosity of the Sun with
time is similar to moving a planet through the HZ, and so the same CO$_2$
weathering process responsible for maintaining habitability on the Earth
through time, the carbonate-silicate weathering cycle, may similarly stabilize
planetary climates within the HZ.

In the carbonate-silicate cycle, atmospheric CO$_2$ dissolves in water and
weathers silicates on both the continents and seafloor, which releases cations
and anions \cite{ebelmen_sur_1845, walker_negative_1981,
    walker_biogeochemical_1993, berner_phanerozoic_2004, mills_changing_2014,
krissansen-totton_constraining_2017,hakim_lithologic_2020}.  On the continents,
the weathering products, including dissolved SiO$_2$, HCO$_{3}^{-}$, and
Ca$^{++}$, wash into the oceans where the HCO$_{3}^{-}$ combine with cations
like Ca$^{++}$ to create CaCO$_3$, which precipitates out of solution. The net
process converts atmospheric CO$_2$ into marine carbonate minerals (i.e.,
CaCO$_3$). Also, seafloor weathering occurs when seawater releases Ca$^{++}$
ions from the seafloor basalt and CaCO$_3$ precipitates in pores and veins.
Subsequently, the carbonates within sediments and altered seafloor can be
subducted. 

Carbon returns to the atmosphere from outgassing. If CO$_2$
outgassing increases above the steady-state outgassing rate, a planet's surface
temperature rises. This leads to increased rainfall and continental
weathering as well as potentially warmer deep-sea temperatures and more
seafloor weathering
\cite{krissansen-totton_constraining_2018,berner_phanerozoic_2004,coogan_alteration_2015}.
Increased weathering draws down atmospheric CO$_2$ and stabilizes the climate
over ${\sim}10^{6}$-year timescales on habitable, Earth-like planets
\cite{kadoya_conditions_2014}. 

One- and three-dimensional climate calculations of HZ limits
\cite{kasting_earths_1993, kopparapu_habitable_2013, yang_strong_2014} assume
that a carbonate-silicate weathering cycle is functioning but do not explicitly
include it. The assumed presence of the carbonate-silicate cycle would predict
that atmospheric CO$_2$ of Earth-like planets increases with orbital distance
in the HZ
\cite{kasting_earths_1993,kopparapu_habitable_2013,kadoya_conditions_2014}. In
particular, future telescopic observations, e.g. NASA's
    Habitable Exoplanet Imaging Mission (HabEx)\cite{habex_habitable_2019} and
    Large Ultraviolet Optical Infrared Surveyor
(LUVOIR)\cite{luvoir_large_2019}, could search for the CO$_2$ trend to test
the HZ hypothesis
\cite{bean_statistical_2017,checlair_statistical_2019,turbet_two_2019}.
Previous studies \cite{abbot_indication_2012, kadoya_conditions_2014} have
suggested the carbonate-silicate weathering cycle could alter predictions of
pCO$_2$ in the HZ, but it is important to know the exact relationship we are
looking for. Also, while an increase of pCO$_2$ with orbital distance in the HZ
may be true if all Earth-like exoplanets have the exact same properties as the
modern Earth, the trend becomes less certain if HZ planetary characteristics
deviate from those of the modern Earth. There could be considerable variability
in atmospheric CO$_2$ throughout the HZ, perhaps even enough to obscure a
monotonic trend with orbital distance.

Here, we show that uncertain physicochemical and geophysical
parameters in the carbonate-silicate weathering cycle
\cite{krissansen-totton_constraining_2017} cause scatter in pCO$_2$ with
orbital distance. We then demonstrate that future telescopes
must observe at least 83 ($2\sigma$) HZ planetary atmospheres to confidently
detect our predicted relationship between atmospheric CO$_2$ and orbital
distance, and confirm the HZ hypothesis.

\section{Results} 

\subsection{Stable pCO$_2$ abundances from our numerical model}

\begin{table}[tbh!]
\centering
\small
\begin{tabular}{||c|p{7cm}|c|c|c||}
 \hline
 Parameter & Parameter Description & Range & Scaling & Units\\
 \hline\hline
 $F_{\textrm{out}}^{\textrm{mod}}$ & Modern CO$_{2}$ outgassing flux & 6-10 & & Tmol C yr$^{-1}$ \\
$n$ & Carbonate precipitation coefficient & 1-2.5 & $\propto \left[\textrm{CO}^{2-}_{3} \right]^{\textrm{n}}$ & \\ 
$x$ & Modern seafloor dissolution relative to precipitation & 0.5-1.5 & $\propto xF_{\textrm{out}}^{\textrm{mod}}$ &\\
$T_{\textrm{e}}$ & E-folding temperature factor for continental weathering & 10-40 & & K \\
$\alpha$ & Power law exponent for CO$_{2}$ dependence of continental silicate weathering & 0.1-0.5 & $\propto \left( \textrm{pCO}_{2} \right)^{\alpha}$ & \\
$\xi$ & Power law exponent for CO$_{2}$ dependence of continental carbonate weathering & 0.1-0.5 & $\propto \left( \textrm{pCO}_{2} \right)^{\xi}$ & \\
$f_{\textrm{land}}$ & Land fraction compared to modern Earth & 0-1 & & \\
$S_{\textrm{thick}}$ & Ocean sediment thickness relative to modern Earth & 0.2-1 & &\\
$F_{\textrm{carb}}^{\textrm{mod}}$ & Modern continental carbonate weathering & 7-14 & & Tmol C yr$^{-1}$ \\
$f_{\textrm{bio}}$ & Biological weathering compared to modern Earth & 0-1 & &\\
$a_{\textrm{grad}}$ & Surface to deep ocean temperature gradient scaling & 0.8-1.4 & $\propto a_{\textrm{grad}}T_{\textrm{s}}$ & \\
$\gamma$ & Power law exponent for pH dependence of seafloor dissolution & 0-0.5 & $\propto \left(\left[\textrm{H}^{+}\right]\right)^{\gamma}$ & \\
$\beta$ & Power law exponent for seafloor spreading rate & 0-0.2 & $\propto Q^{\beta}$ &\\
$m$ & Exponent for outgassing dependence on crustal production & 1-2 & $\propto Q^{\textrm{m}}$ & \\
$E_{\textrm{bas}}$ & Seafloor dissolution activation energy & 60-100 & & kJ mol$^{-1}$ \\
$n_{\textrm{out}}$ & Exponent for internal heat with time & 0-0.73 & see eq. \ref{internal_heat_eqn} & \\
$\tau$ & Planet age (see eq. \ref{internal_heat_eqn})* & 0-10 & & Gyr \\
$S$ & Incident flux relative to modern Earth* & 0.35-1.05 & & \\
\hline
\end{tabular}
\caption{Parameter ranges for our numerical model. Parameters are dimensionless
    unless otherwise described. The fourth column shows how scaling parameters
    impact the model, where $T_{\textrm{s}}$ is the surface temperature in K and $Q$ is
    the internal heat of the planet relative to the modern Earth (see equation
    \ref{internal_heat_eqn} for $Q$). Unless otherwise
    noted, each parameter range is justified in the original
    model derivation for the Earth through
    time\cite{krissansen-totton_constraining_2018}. *The justification for
    this parameter is given in the Methods, subsection
    \nameref{sec:numerical_model}.
}
\label{tab:model_params}
\end{table}

We use a coupled climate and carbonate-silicate weathering model (see
Methods, subsection \nameref{sec:numerical_model}) to explore
pCO$_2$ on Earth-like planets in the HZ. The model considers numerous planetary
properties, listed in Table \ref{tab:model_params}, and their effect on the
carbonate-silicate weathering cycle to calculate a planet's steady-state
pCO$_2$ and surface temperature. If the globally averaged, steady-state surface
temperature is below 248 K, we assume the planet is completely frozen and
uninhabitable at the surface, as shown by three-dimensional climate models
\cite{charnay_exploring_2013}. Similarly, we assume planets are uninhabitable
beyond 355 K, above which surface water would be rapidly lost to space
\cite{wolf_constraints_2017} (see Methods, subsection
\nameref{sec:numerical_model} for additional details on these assumed
temperature constraints).

\begin{figure}[htb!]
\centering
\includegraphics[width=10cm]{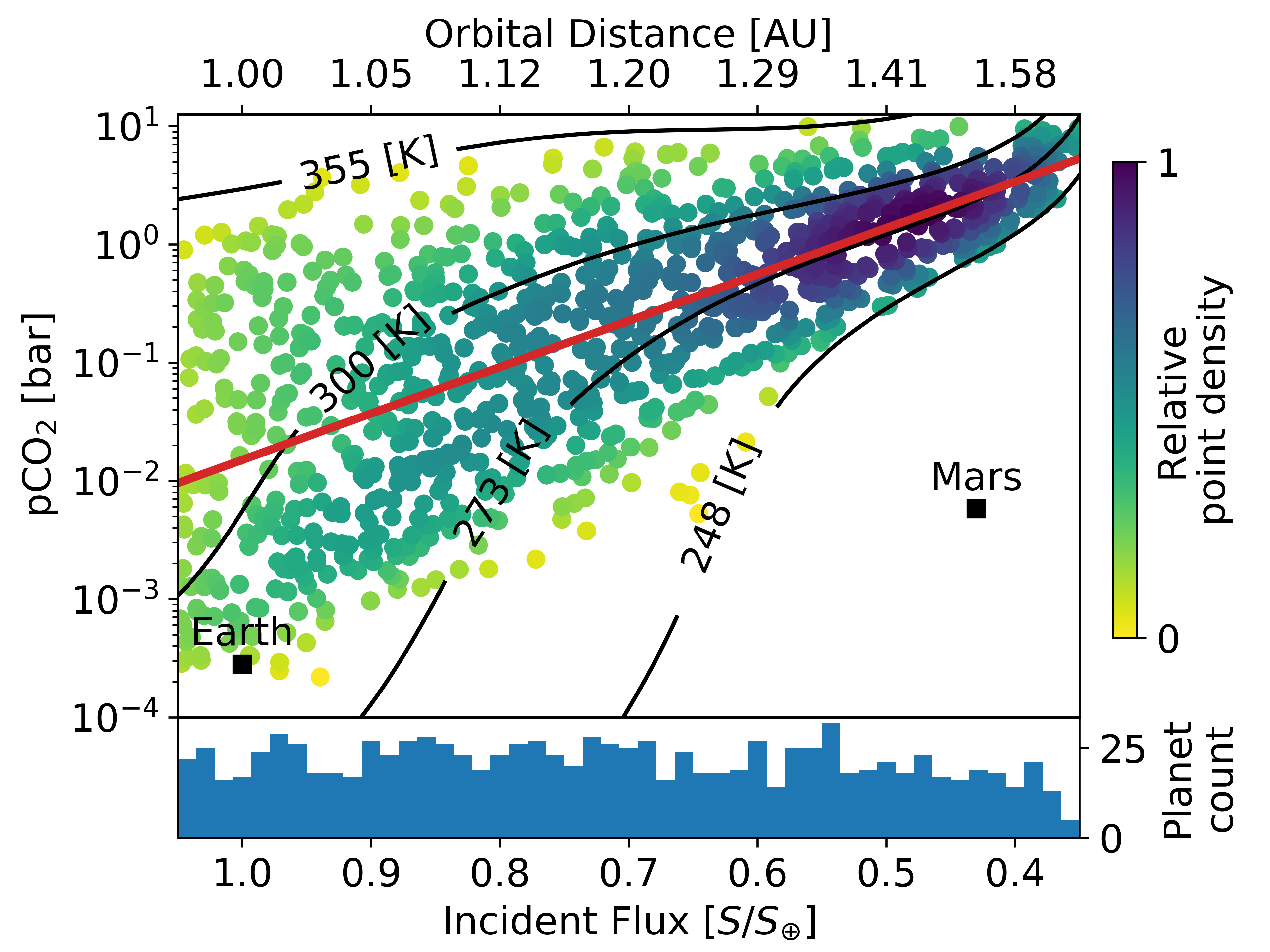}
\caption{The expected distribution of stable, Earth-like exoplanet climates
from our habitable zone weathering model. The horizontal axis shows incident
flux, $S$, normalized to the solar constant ($S_{\oplus}$) and the
corresponding orbital distance in Astronomical Units (AU) above the plot. The
vertical axis shows the atmospheric CO$_2$ partial pressure (pCO$_2$) in bar.
Each point represents a climate in steady state. The black labeled contours
show the mean global surface temperature for the given pCO$_2$ and incident
flux. The white region below the 248 K contour is where our model assumption of
a liquid ocean is no longer plausible so no planets are shown in that region.
Above the 355 K contour Earth-like planets are too hot to retain their liquid
oceans for billions of years. Similar to the frozen planets, such hot planets
are not considered habitable. Modern Earth and Mars are shown by black squares.
The blue histogram at the bottom of the figure shows the number of stable
planets in each incident flux bin. The color of each simulated planet shows the
relative point density in the plot at that location. The color was calculated
using a kernel-density estimate with Gaussian kernels and rescaled from 0 to 1.
A color value of 0 represents the lowest relative point density, 1 the highest.
The log-linear line of best fit between pCO$_2$ and $S$ is shown in red. The
slope of the red, best fit line is 3.92$\pm$0.24 (95\%) with units
-$\log_{10}$(pCO$_{2}$ [bar])/[$S/S_{\oplus}$]. Our model predicts that
atmospheric CO$_2$ should increase with orbital distance in the HZ.}
\label{fig:sim_exo_data}
\end{figure}

We randomly generated 1050 habitable, stable, Earth-like exoplanet climates
using uniform distributions of the model parameters in Table
\ref{tab:model_params}. A total of 1200 random, initial parameter combinations
were considered but we eliminated those that resulted in planets that froze
completely or were too hot to retain their surface oceans. As colored dots,
Figure \ref{fig:sim_exo_data} shows habitable, steady-state solutions.

Our model predicts that atmospheric CO$_2$ abundances should broadly increase
and narrow in their spread with orbital distance in the HZ (Figure
\ref{fig:sim_exo_data}), consistent with other models of CO$_2$ in the HZ
\cite{kadoya_conditions_2014, graham_thermodynamic_2020}. As justified next in
Section \ref{sec:theory}, the scatter is about a nominal linear trend between
incident flux, $S$, and log(pCO$_2$), which is different from a non-linear
trend in models that assume a constant surface temperature in the HZ from
negative feedbacks \cite{bean_statistical_2017,turbet_two_2019} but do not
actually model the carbonate-silicate feedbacks. If future
missions are to test the HZ concept by searching for a trend between incident
flux, $S$, and pCO$_{2}$ \cite{bean_statistical_2017,
checlair_statistical_2019, turbet_two_2019}, they could search for the
fundamental $S$-pCO$_2$ relationship shown in Figure \ref{fig:sim_exo_data}. 

Below, we show that a log-linear relationship between pCO$_2$ and $S$ may be
the default in the HZ if Earth-like carbonate-silicate weathering is ubiquitous
on habitable planets. In fact, the trend is elucidated by combining climate
theory with carbonate-silicate cycle theory in what follows.

\subsection{Habitable zone climate theory revisited} \label{sec:theory}

\begin{figure}[htb!]
\centering
\includegraphics[width=10cm]{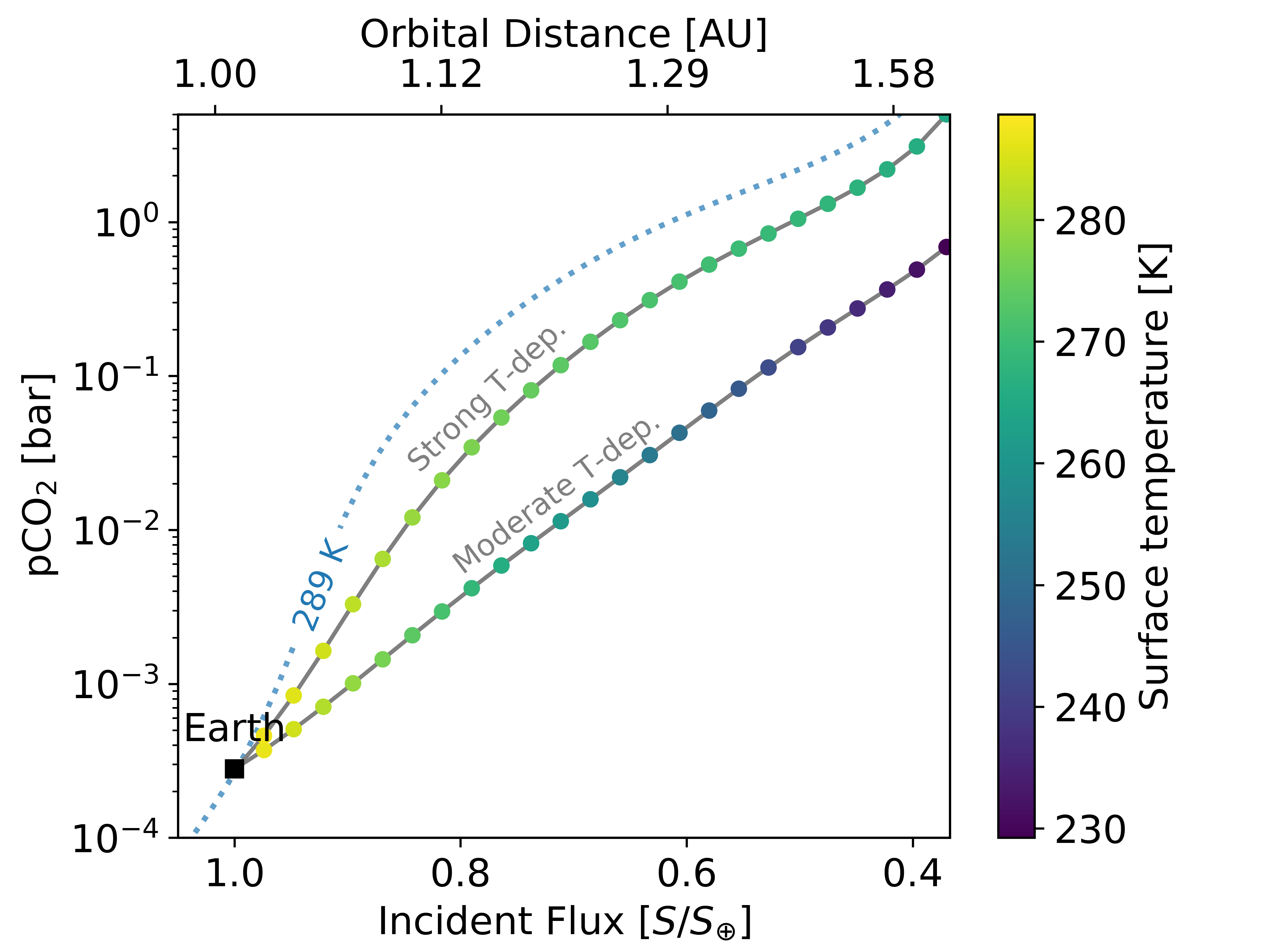}
\caption{The relationship between incident flux and atmospheric CO$_2$ for
Earth-like planets regulated by a carbonate-silicate weathering cycle. The
horizontal axis shows incident flux, $S$, normalized to the solar constant
($S_{\oplus}$) and the corresponding orbital distance in Astronomical Units
(AU) above the plot. The vertical axis shows the atmospheric CO$_2$ partial
pressure (pCO$_2$) in bar. The dotted blue curve labelled 289 K shows the
pCO$_2$ value required to maintain a 289 K surface temperature for the
given incident flux, $S$. The conventional assumption of CO$_2$ in the HZ
stipulates that pCO$_2$ will adjust to maintain a temperate or even
constant surface temperature. Under this assumption, moving the modern
Earth (labelled black square) outward in the HZ would have the planet
approximately follow the dotted blue 289 K contour. The colored points and
gray curves show the modern Earth moving outward in the HZ with a
functioning carbonate-silicate weathering cycle, calculated from equation
\ref{eqn:T_s_carb_sil}. We consider two temperature and pCO$_2$
dependencies for continental weathering in this plot.  The strong
temperature dependence contour (labelled Strong T-dep.), uses a temperature
and pCO$_2$-dependent weathering factor of $\alpha T_{\textrm{e}}= 2.3$,
which implies a strong temperature feedback on continental weathering
compared to the pCO$_2$ feedback (see equation \ref{eqn:S_pco2_final}). The
moderate temperature dependence contour (labelled Moderate T-dep.), uses a
temperature and pCO$_2$-dependent weathering factor of $\alpha
T_{\textrm{e}}=7.5$. These two values for $\alpha T_{\textrm{e}}$ result in
two different paths the Earth can take as it moves outward in the HZ. The
planet color shows the mean surface temperature. Log-linear fits to the
colored points of the Strong T-dep. and Moderate T-dep. contours have
r$^{2}$ values of 0.959 and 0.999, respectively. Thus, even for a strong
temperature dependence of continental weathering, our coupled climate and
weathering model predicts an approximately log-linear relationship between
incident flux and pCO$_2$ on Earth-like planets in the HZ.}
\label{fig:linear_fit}
\end{figure}

A conventional assumption is that the carbonate-silicate weathering cycle will
approximately maintain a stable, temperate surface temperature for an
Earth-like planet moved about in the HZ \cite{kasting_climate_1989,
kasting_habitable_1993, kopparapu_habitable_2013} or even a constant
temperature \cite{bean_statistical_2017, turbet_two_2019}. Thus, if we moved
the modern Earth outward in the HZ, the smaller incident flux would initially
cause the planet to cool. The cooler temperature would lower the CO$_2$
weathering rate causing CO$_2$ to accumulate in the atmosphere until the
temperature returned to its nominal value of 289 K. Figure \ref{fig:linear_fit}
shows this scenario with the dotted blue 289 K contour, which gives the pCO$_2$
value required to maintain a 289 K surface temperature for the modern Earth as
it moves about the HZ. The line was calculated from a radiative-convective
climate model described in the Methods below, subsection
\nameref{sec:expanded_climate_model} (see equation
\ref{eqn:climate_fit_eqn}).

The constant, 289 K surface temperature contour in Figure \ref{fig:linear_fit}
is a non-linear relationship between incident flux, $S$, and log(pCO$_2$) but
it does not consider the temperature and pCO$_2$ feedbacks inherent to the
carbonate-silicate weathering cycle. We demonstrate that if these feedbacks are
taken into account, surface temperature declines with orbital distance, as
mentioned in previous work \cite{kadoya_conditions_2014}, and
the relationship between $S$ and log(pCO$_2$) is actually approximately linear
for Earth-like planets in the HZ.

If Bond albedo is fixed, the surface temperature, $T_{\textrm{s}}$, for an Earth-like
planet in steady-state varies approximately linearly with incident flux, $S$
\cite{budyko_effect_1969,koll_earths_2018,catling_atmospheric_2017}. This
linear relationship between $T_{\textrm{s}}$ and $S$ arises from energy balance and from
water vapor feedback and can be expressed as

\begin{equation}\label{eqn:flux_and_temp}
    F_{\textrm{SOL}} = F_{\textrm{OLR}} = \left( \frac{1-A_{\textrm{B}}}{4} \right)S = a + bT_{\textrm{s}}
\end{equation}
where $F_{\textrm{SOL}}$ is the incoming solar radiation flux, $F_{\textrm{OLR}}$ is the outgoing
long-wavelength radiation flux, $A_{\textrm{B}}$ is the Bond albedo, and $a$ and $b$ are
empirical constants. From satellite measurements of the modern Earth and
radiative calculations, for $T_{\textrm{s}}$ in K, the empirical constants in equation
\ref{eqn:flux_and_temp} are approximately $a=-370$ W${\cdot}$m$^{-2}$ and
$b=2.2$ W${\cdot}$m$^{-2}{\cdot}$K$^{-1}$ \cite{koll_earths_2018}. 

Solving for $T_{\textrm{s}}$ in equation \ref{eqn:flux_and_temp}, the surface temperature
is given by

\begin{equation}\label{eqn:T_s}
    T_{\textrm{s}} = \left( \frac{1-A_{\textrm{B}}}{4b} \right)S - \frac{a}{b}.
\end{equation}
Under the conventional assumption that the HZ is regulated by a CO$_2$-H$_2$O
greenhouse effect where H$_2$O concentrations respond to changes in pCO$_2$,
the temperature offset in equation \ref{eqn:T_s}, $-a/b$, is a function of
pCO$_2$. Thus, surface temperature, as a function of $S$ and pCO$_2$, is given
by

\begin{equation}\label{eqn:T_s_with_fpco2}
    T_{\textrm{s}}\left(S,\textrm{pCO}_{2}\right) = \left( \frac{1-A_{\textrm{B}}}{4b} \right)S + f\left( \textrm{pCO}_{2} \right).
\end{equation}
where $f\left( \textrm{pCO}_{2}\right)$ is a function that depends on pCO$_2$.
For the modern Earth at 1 AU, $f(\textrm{pCO}_{2})=-a/b$. For pCO$_{2}\leq0.1$
bar, the CO$_2$ greenhouse effect is logarithmic in pCO$_2$, i.e.,
$f(\textrm{pCO}_{2})\propto \ln{(\textrm{pCO}_{2})}$
\cite{myhre_new_1998,pierrehumbert_principles_2010}. Above ${\sim}0.1$ bar,
weaker CO$_2$ absorption features become important and $f(\textrm{pCO}_{2})$
deviates from $\propto \ln{(\textrm{pCO}_{2})}$
\cite{charnay_warm_2017,pierrehumbert_principles_2010}. 

As pCO$_2$ increases for an Earth-like planet moved outward in the HZ, the
surface temperature will follow equation \ref{eqn:T_s_with_fpco2} while the
rate at which CO$_2$ is removed from the atmosphere will adjust according to
the carbonate-silicate weathering feedback. Quantitatively, the pCO$_2$- and
$T_{\textrm{s}}$-dependent flux of CO$_2$ removal due to the continental weathering
flux, $F_{\textrm{w}}$ (in mol CO$_2$ per unit time) is described by

\begin{equation}\label{eqn:weathering_temp_dependence_full}
    F_{\textrm{w}} = \rho
    \left(\frac{\textrm{pCO}_{2}}{\textrm{pCO}_{2}^{\textrm{mod}}}\right)^{\alpha} \exp{\left(\frac{T_{\textrm{s}}\left(S,\textrm{pCO}_{2}\right)-T_{\textrm{s}}^{\textrm{mod}}}{T_{\textrm{e}}}\right)}
\end{equation}
where $\rho$ is a constant determined by the continental weathering properties
of the modern Earth, $\alpha$ is a dimensionless constant between 0.1 and 0.5
and regulates the pCO$_2$ dependence of continental silicate
weathering, $T_{\textrm{e}}$ is a constant between 10 K and 40 K and represents the
e-folding temperature dependence of continental weathering. The range for
$T_{\textrm{e}}$ has been empirically constrained for the surface temperatures relevant
to habitable, Earth-like planets from lab measurements and Phanerozoic geologic
constraints \cite{walker_negative_1981, krissansen-totton_constraining_2017}.
Finally, pCO$^{\textrm{mod}}_{2}=288\times10^{-6}$ bar and
$T_{\textrm{s}}^{\textrm{mod}}=289$ K are the modern Earth's preindustrial pCO$_2$ and
surface temperature, respectively \cite{krissansen-totton_constraining_2018}. 

The range for $\alpha$ on the Earth was determined empirically from geologic
constraints over the past 100 Myr \cite{krissansen-totton_constraining_2017}.
We assume that this derived range for $\alpha$ applies to the Earth through
time \cite{krissansen-totton_constraining_2018, krissansen-totton_coupled_2020,
kadoya_probable_2020} and the Earth-like exoplanets modelled here that have a
carbonate-silicate cycle. However, better proxy data for the ancient Earth or
observing the carbonate-silicate cycle on habitable exoplanets
\cite{bean_statistical_2017, turbet_two_2019} may be necessary to understand if
the assumed range for $\alpha$ applies more generally to habitable planets. 

In equation \ref{eqn:weathering_temp_dependence_full}, we assume seafloor
weathering is negligible, which is a reasonable approximation for the modern
Earth \cite{krissansen-totton_constraining_2018}, and illustrative for our
purposes of deriving a simple, analytic relationship between $S$ and pCO$_2$.
Here, we seek to predict the behavior of the modern Earth in the HZ to gain
intuitive understanding, whereas in our numerical model we consider a broad
range of properties for Earth-like planets on which seafloor weathering may be
important.

The modern Earth, and all Earth-like planets considered in this work, are
assumed to be in steady-state, in which the flux of CO$_2$ from volcanic
outgassing is equal to the rate of CO$_2$ removal from weathering, $F_{\textrm{w}}$. If
we assume a HZ planet with CO$_2$ outgassing the same as the modern Earth's,
$F_{\textrm{w}}$ remains constant despite changes in $S$ and pCO$_2$. Setting
$T_{\textrm{s}}\left(S,\textrm{pCO}_{2}\right)=T_{\textrm{s}}^{\textrm{mod}}$ and
$\textrm{pCO}_{2}=\textrm{pCO}_{2}^{\textrm{mod}}$ for the modern Earth, from
equation \ref{eqn:weathering_temp_dependence_full}, we see that $F_{\textrm{w}}=\rho$
and 

\begin{equation}\label{eqn:carb_sil_eq_1}
    1 =
    \left(\frac{\textrm{pCO}_{2}}{\textrm{pCO}_{2}^{\textrm{mod}}}\right)^{\alpha} \exp{\left(\frac{T_{\textrm{s}}\left(S,\textrm{pCO}_{2}\right)-T_{\textrm{s}}^{\textrm{mod}}}{T_{\textrm{e}}}\right)}.
\end{equation}
Equation \ref{eqn:carb_sil_eq_1} must hold for a modern Earth within the
HZ. If it did not, $F_{\textrm{w}}$ would not balance CO$_2$ outgassing, which would
result in either complete CO$_2$ removal, or CO$_2$ accumulation without bound. 

Solving for $T_{\textrm{s}}\left(S,\textrm{pCO}_{2}\right)$ in equation
\ref{eqn:carb_sil_eq_1} we find

\begin{equation}\label{eqn:T_s_carb_sil}
    T_{\textrm{s}}\left(S,\textrm{pCO}_{2}\right) = T_{\textrm{s}}^{\textrm{mod}} - \alpha T_{\textrm{e}} \ln{\left(\frac{\textrm{pCO}_{2}}{\textrm{pCO}_{2}^{\textrm{mod}}}\right)}. 
\end{equation}
Equating equation \ref{eqn:T_s_carb_sil} to equation \ref{eqn:T_s_with_fpco2}
and rearranging gives

\begin{equation}\label{eqn:S_pco2_final}
    \left( \frac{1-A_{\textrm{B}}}{4b} \right)S = T_{\textrm{s}}^{\textrm{mod}} - \left[ \alpha T_{\textrm{e}} \ln{\left(\frac{\textrm{pCO}_{2}}{\textrm{pCO}_{2}^{\textrm{mod}}}\right)} + f\left(\textrm{pCO}_{2}\right) \right]. 
\end{equation}
If $f(\textrm{pCO}_{2})\propto \ln{(\textrm{pCO}_{2})}$, which is the case for
pCO$_{2} \leq 0.1$ bar \cite{charnay_warm_2017,pierrehumbert_principles_2010},
then $S\propto -\ln{(\textrm{pCO}_{2})}$. However, even if
$f(\textrm{pCO}_{2})$ deviates from log-linearity with pCO$_2$, $S$ will become
increasingly log-linear with pCO$_2$ as $\alpha T_{\textrm{e}}$ increases. In equation
\ref{eqn:S_pco2_final}, increasing $\alpha T_{\textrm{e}}$ will cause the
$\ln{(\textrm{pCO}_{2})}$ term to dominate the relationship between $S$ and
pCO$_2$. Intuitively, increasing $\alpha T_{\textrm{e}}$ decreases the temperature dependence
of continental weathering relative to its pCO$_2$ dependence. Note that bigger
$T_{\textrm{e}}$ reduces the temperature dependence of continental weathering while
bigger $\alpha$ increases the pCO$_2$ dependence of continental weathering (equation
\ref{eqn:weathering_temp_dependence_full}). 

In addition to predicting a linear relationship between log(pCO$_2$) and $S$,
the carbonate-silicate cycle implies that moving an Earth-like planet outward
in the HZ will cause $T_{\textrm{s}}\left(S,\textrm{pCO}_{2}\right)$ to decrease. For
increasing orbital distance, pCO$_2$ must increase for
$T_{\textrm{s}}\left(S,\textrm{pCO}_{2}\right)$ to increase in the HZ. From equation
\ref{eqn:carb_sil_eq_1}, pCO$_2$ will be greater than pCO$_{2}^{\textrm{mod}}$
in such cases so $T_{\textrm{s}}\left(S,\textrm{pCO}_{2}\right)$ must be less than
$T_{\textrm{s}}^{\textrm{mod}}$. This decrease in $T_{\textrm{s}}\left(S,\textrm{pCO}_{2}\right)$
as $S$ decreases is shown in Figure \ref{fig:linear_fit}. Physically, the power
law dependence of weathering on pCO$_2$ can balance volcanic outgassing at
lower surface temperatures in the outer HZ.

Figure \ref{fig:linear_fit} shows the approximately log-linear relationship
between pCO$_2$ and $S$ for the modern Earth moved outward in the HZ. The gray
lines and colored circles in Figure \ref{fig:linear_fit} show the expected
pCO$_2$ value for the given incident flux $S$, calculated from equation
\ref{eqn:T_s_carb_sil}. For each $S$ value in Figure \ref{fig:linear_fit},
equation \ref{eqn:T_s_carb_sil} was solved for pCO$_2$ by using equation
\ref{eqn:climate_fit_eqn}, the polynomial fit for surface temperature based on
a 1D climate model (described in the Methods, subsection
\nameref{sec:expanded_climate_model}), assuming values of
$\alpha T_{\textrm{e}}$.

The value of $\alpha T_{\textrm{e}}$ affects the slope of the relationship between $S$ and
pCO$_2$ due to the carbonate-silicate weathering cycle, shown in Figure
\ref{fig:linear_fit}. From above, the ranges for $\alpha$ and $T_{\textrm{e}}$ are
$0.1\leq \alpha \leq 0.5$ and $10 \leq T_{\textrm{e}} \leq 40$
\cite{krissansen-totton_constraining_2018}, so $1 \leq \alpha T_{\textrm{e}} \leq 20$. If we
consider uniform distributions of $\alpha$ and $T_{\textrm{e}}$ then roughly 95\% of
$\alpha T_{\textrm{e}}$
values will be greater than 2.3. If $\alpha=0.23$ and $T_{\textrm{e}}=10$ K then
$\alpha T_{\textrm{e}}=2.3$, which is used for the Strong T-dep. curve in Figure
\ref{fig:linear_fit}. The mean of each parameter, $\alpha=0.3$ and $T_{\textrm{e}}=25$ K
gives $\alpha T_{\textrm{e}}=7.5$, which corresponds to the Moderate T-dep. curve in Figure
\ref{fig:linear_fit}. For $\alpha T_{\textrm{e}}\leq 2.3$ the colored points and gray curves
become increasingly similar to the constant 289 K surface temperature contour
in Figure \ref{fig:linear_fit}. However, for uniform distributions of $\alpha$ and
$T_{\textrm{e}}$, ${\sim}$95\% of $\alpha T_{\textrm{e}}$ values are greater than 2.3 so an approximately
log-linear relationship between $S$ and $\log{(\textrm{pCO}_{2})}$ is the
default expectation for Earth-like planets in the HZ.

\subsection{Observational uncertainty for pCO$_2$ in the HZ}
\label{sec:results}

In the log-linear fit shown as the solid red line in Figure
\ref{fig:sim_exo_data}, which is the expected relationship between pCO$_2$ and
$S$ that we have derived above, the r$^2$-value is 0.49. Thus, about half the
variance in log(pCO$_2$) is described by changes in incident flux. The slope is
3.92$\pm$0.24 (95\%) with units -$\log_{10}$(pCO$_{2}$ [bar])/($S/S_{\oplus}$)
so our model predicts a trend of increasing atmospheric CO$_2$ with orbital
distance, which future missions might detect \cite{bean_statistical_2017,
turbet_two_2019, checlair_statistical_2019}. However, there is
sufficient spread in our simulated planets that confirming the HZ hypothesis
from such a trend may be difficult. 

This difficulty is readily seen if we consider a log-uniform
distribution for pCO$_2$ on Earth-like planets in the HZ. If we randomly
generate 1050 such planets, where ${10^{-4}\leq}$pCO$_{2}{\leq 10}$ bar is
sampled log-uniformly, $0.35 \leq S \leq 1.05$ is sampled uniformly, and impose the same
constraints on surface temperature for habitability as in Figure
\ref{fig:sim_exo_data}, then the log-linear line of best fit through the
uniform planet data has a slope of $3.76 \pm 0.465$ (95\%) with units
-$\log_{10}$(pCO$_{2}$ [bar])/($S/S_{\oplus}$). Thus, measuring just the
log-linear trend between pCO$_2$ and $S$ in the HZ is unlikely to test the HZ
hypothesis as this measurement cannot confidently detect the presence of the
carbonate-silicate weathering cycle \textemdash{} it is indistinguishable from
that of randomly distributed pCO$_2$ between the surface temperature limits for
habitability.

The inability to differentiate between the log-linear trends
for weathering-mediated and random pCO$_2$ vs $S$ in the HZ is due to the
assumed surface temperature constraints we impose in our model (between 248
and 355 K for planets in the HZ, see Methods, subsection
\nameref{sec:numerical_model}). Such
temperature constraints are necessary as the carbonate-silicate weathering
cycle can only operate when water, as liquid, is present at the planetary
surface. Even without the carbonate-silicate weathering cycle, a minimum
surface temperature for habitable planets, which must exist, will result in
increasing pCO$_2$ with orbital distance, as shown by the constant temperature
contours in Figure \ref{fig:sim_exo_data}.

To test the HZ hypothesis, we propose searching for the
two-dimensional distribution of planets in the $S$-pCO$_2$ phase space that
arises from the carbonate-silicate weathering cycle. This
$S$-log(pCO$_2$) relationship is shown by the point density in Figure
\ref{fig:sim_exo_data}, where the distribution of habitable, stable planets is
not log-uniformly distributed over pCO$_2$. Rather, around the best-fit line,
there is an abundance of planets in the outer HZ at high pCO$_2$, a dearth of
low pCO$_2$ planets between ${\sim}$0.9 and ${\sim}$0.7 $S/S_{\oplus}$, and few
high-pCO$_2$ planets throughout the HZ compared to the log-uniform pCO$_2$
case. These differences are expected features of the carbonate-silicate
weathering cycle due to the temperature- and pCO$_2$-dependent nature of the
weathering feedback. Recall from Section \ref{sec:theory} that, as $S$
decreases, the lowered temperature will reduce weathering causing pCO$_2$ to
increase. This results in the lack of low-pCO$_2$ planets in the middle of the
HZ and the high abundance of habitable planets in the outer HZ (purple shaded
region in Figure \ref{fig:sim_exo_data}).  Similarly, for large pCO$_2$, the
temperature is warmer and pCO$_2$ higher than that of modern Earth so the
carbonate-silicate weathering cycle acts to lower pCO$_2$, which reduces the
number of high-pCO$_2$ planets throughout the HZ relative to the outer HZ.

To detect the prevalence of the carbonate-silicate weathering
cycle and test the validity of the HZ concept, future observations should
measure the two-dimensional $S$-pCO$_2$ distribution of habitable, Earth-like
exoplanets.  This measured distribution can be compared to the distribution
of habitable planets we predict in Figure \ref{fig:sim_exo_data} to determine
if Earth-like planets in the HZ are consistent with the $S$-pCO$_2$ predictions
of the carbonate-silicate weathering cycle.

\begin{figure}[htb!]
\centering
\includegraphics[width=10cm]{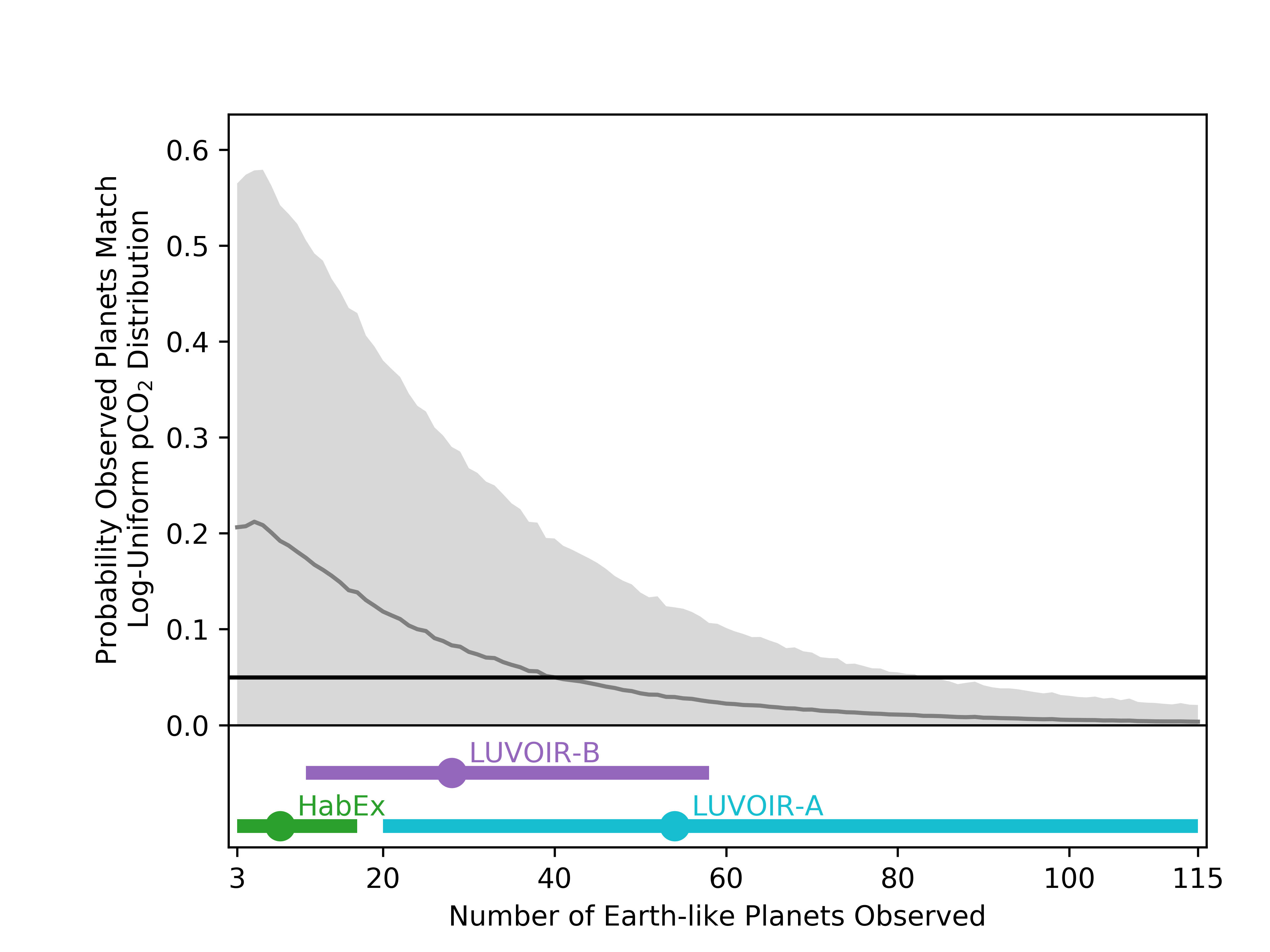}
\caption{The probability observed exoplanets will accidentally match a
log-uniform distribution for pCO$_2$ in the HZ if the true pCO$_2$
distribution is regulated by the carbonate-silicate weathering cycle, as
shown in Figure \ref{fig:sim_exo_data}. This probability is shown on the
vertical axis. The horizontal axis shows the number of observed Earth-like
exoplanets. The solid gray curve and corresponding shaded gray region shows
the mean probability and $2\sigma$ uncertainty, respectively, that the
observed planets, sampled from the planets shown in Figure
\ref{fig:sim_exo_data}, match a log-uniform pCO$_2$ distribution in the HZ.
This curve is calculated from 10,000 two-dimensional Kolmogorov-Smirnov
tests (see Results, subsection \nameref{sec:results}).
The solid, horizontal black line highlights the 5\% probability line. At
the bottom of the figure, the labeled points and error bars show the number
of Earth-like exoplanets the next generation of proposed space telescopes
are expected to observe (telescope data in Table \ref{tab:model_params}).
The vertical scaling of the telescope points is arbitrary, only the
horizontal position and extent of the $1{\sigma}$ error bars is
significant. To rule out a log-uniform pCO$_2$ distribution with 95\%
confidence, future telescopes would need to observe at least 83 Earth-like
planets. \label{fig:ks_trend}}
\end{figure}

A test of the two-dimensional phase space of $S$ and pCO$_2$
in the HZ is shown in Figure \ref{fig:ks_trend}, which was produced using a
two-dimensional Kolmogorov-Smirnov (KS) test. The two-dimensional KS test
compares the statistical similarity of a sample distribution to a reference
distribution
\cite{peacock_two-dimensional_1983,fasano_multidimensional_1987,press_kolmogorov-smirnov_1988}.
For Figure \ref{fig:ks_trend}, the reference distribution was comprised of 500
randomly generated planets from the log-uniform distribution for pCO$_2$
described above ($10^{-4}\leq$pCO$_2 \leq 10$ bar, $0.35\leq S \leq 1.05$
$S/S_{\oplus}$, and surface temperature between 248 and 355 K). The sample
distribution was generated by randomly selecting a number of planets from
Figure \ref{fig:sim_exo_data} equal to the number of
observed exoplanets. For a given number of observed exoplanets in Figure
\ref{fig:ks_trend}, the horizontal axis, we ran the KS test 10,000 times then
calculated the mean and standard deviation from those runs, shown by the gray
contour and shaded region. This resampling is necessary as the two-dimensional
KS test is a nonparametric approximation that two data sets come from the same
underlying population \cite{fasano_multidimensional_1987}.  We note that below
${\sim}$20 observed planets and for probabilities above ${\sim}$0.1, the
two-dimensional KS test used here can be unreliable
\cite{press_kolmogorov-smirnov_1988}. These limitations do not invalidate the
analysis shown in Figure \ref{fig:ks_trend}, as we want to know, with 95\%
confidence, that a log-uniform pCO$_2$ distribution can be ruled out if real
exoplanets follow the distribution shown in Figure \ref{fig:sim_exo_data},
which corresponds to the gray line and shaded contour dipping below the 0.05
probability value, shown by the horizontal black line, at 83 observations in
Figure \ref{fig:ks_trend}.

\begin{table}[htb!]
\centering
\begin{tabular}{||c | c | c ||} 
 \hline
 Telescope & Diameter [m] & Expected Yield ($1{\sigma}$)\\
 \hline\hline
 HabEx & 4 & $8^{+9}_{-5}$ \\
 LUVOIR-B & 8 & $28^{+30}_{-17}$ \\
 LUVOIR-A & 15 & $54^{+61}_{-34}$ \\
 \hline
\end{tabular}
\caption{The number of expected Earth-like exoplanets observed by each platform
from the HabEx \cite{habex_habitable_2019} and LUVOIR \cite{luvoir_large_2019}
final reports.}
\label{tab:exp_eleps}
\end{table}

Thus, confidently detecting the carbonate-silicate weathering cycle will
require many exoplanet observations, as shown in Figure \ref{fig:ks_trend}.
Proposed NASA telescopes, HabEx and LUVOIR, are expected to observe between 3
and 115 Earth-like exoplanets \cite{luvoir_large_2019,habex_habitable_2019}
(see Table \ref{tab:exp_eleps}).  The ranges for each mission concept are shown
by the colored circles with error bars in Figure \ref{fig:ks_trend}. Only the
nominal capability of LUVOIR-A, the variant of the proposed LUVOIR
space telescope with a primary mirror diameter of 15 m, would provide sufficient
Earth-like exoplanet detections to confidently discriminate between a
log-uniform pCO$_2$ distribution in the HZ and a pCO$_2$ distribution regulated
by the carbonate-silicate weathering cycle. A caveat is that this calculation
does not consider the instrument uncertainty in derived pCO$_2$ measurements
for each telescope or that other processes not considered in our model may
alter pCO$_2$ in the HZ, as discussed below.

\section{Discussion} \label{sec:discussion}

Our model assumes that the full variation and uncertainty in Earth's carbon
cycle parameters through time (Table 1) are representative of habitable
Earth-like exoplanets generally. This assumption is a reasonable first-order
approximation as the bulk composition and geochemistry of rocky exoplanets
appear similar to Earth's \cite{doyle_oxygen_2019}. However, the validity of
this assumption likely depends on the parameter in question. For example, it is
probably reasonable to expect habitable exoplanets to have a wide range of land
fractions and outgassing fluxes, but it is unclear whether there is as much
natural variability in the temperature dependence of silicate weathering. An
improved mechanistic understanding of weathering on Earth
\cite{maher_hydrologic_2014, winnick_relationships_2018} might reduce these
uncertainties.

Other weathering feedbacks have been proposed to operate on the Earth through
time, such as reverse weathering \cite{isson_reverse_2018}. In reverse
weathering, cations and dissolved silica released from silicate weathering are
sequestered into clay minerals rather than carbonates so that CO$_2$ remains in
the atmosphere, warming the climate and reducing ocean pH. Reverse weathering
is thought to be strongly pH dependent and as ocean pH decreases, reverse
weathering turns off, acting as a climate stabilization mechanism similar to
the carbonate-silicate cycle. The importance of reverse weathering is so poorly
constrained through Earth's history \cite{krissansen-totton_coupled_2020} that
it does not make sense to consider it in our model. However, with future
constraints from geology and lab measurements, reverse weathering might
alter the stable CO$_2$ abundances of our modeled atmospheres shown in
Figure \ref{fig:sim_exo_data}.

At both the inner and outer edges of the HZ, our model assumes that abundant
liquid water exists at the planetary surface because, without a liquid surface
ocean, the carbonate-silicate weathering cycle ceases and CO$_2$ cannot be
sequestered after outgassing. Beyond these temperature bounds,
other processes must regulate pCO$_2$. This is a caveat to consider in future
observations. As we see from Figure \ref{fig:sim_exo_data}, Mars has low
atmospheric CO$_2$ and low incident flux. Frozen exoplanets similar to Mars,
populating the white area under the 248 K contour in Figure
\ref{fig:sim_exo_data}, could exist in exoplanet surveys. Similarly, planets
devoid of surface water, such as Venus, might exist at
high pCO$_2$ within the HZ. If future observations detect such planets without
confirming the existence of a liquid surface or surface temperature, it could
introduce additional uncertainties in any relation between orbital distance and
atmospheric CO$_2$. Detecting a surface ocean, one of the most important
surface features to confirm when searching for biosignatures and habitability
\cite{lustig-yaeger_detecting_2018, robinson_detecting_2010,
williams_detecting_2008}, is also important to interpret trends of CO$_2$ in
the HZ.

Because we only consider variations on an Earth-like planet, our model
predictions may underestimate the inherent variability in habitable
exoplanetary conditions. Planets very different from the modern Earth, such as
waterworlds without a carbonate-silicate weathering cycle
\cite{kite_habitability_2018} or CH$_4$-rich worlds
\cite{haqq-misra_revised_2008,wordsworth_transient_2017}, could introduce
additional uncertainty in an observed relationship between $S$ and pCO$_2$ in
the HZ. Despite such uncertainties, future missions should measure the
relationship between $S$ and pCO$_2$ in the HZ, or possibly a
sharp transition in pCO$_2$ at the inner edge of the HZ due to loss of surface
water and subsequent shutoff of surface weathering
\cite{turbet_runaway_2019,graham_thermodynamic_2020}. A more complex model than
presented here is necessary to predict such a jump in pCO$_2$ at the inner edge
of the HZ. However, if the carbonate-silicate weathering cycle is indeed
ubiquitous, as is typically assumed in HZ calculations, then the relationship
between incident flux and pCO$_2$ may follow the $S$-pCO$_2$ relationship
predicted in Figure \ref{fig:sim_exo_data}. If no such relationship is
observed, then the carbonate-silicate weathering cycle may have limited
influence on planetary habitability and the limits of the conventional HZ could
need revision. Alternatively, the HZ hypothesis could be incorrect and the
long-term climate of HZ planets could be set by phenomena
beyond those considered here. 

A previous version of this work was published as part of a Ph.D. thesis
\cite{lehmer_formation_2020}.

\section{Methods} \label{sec:methods}

\subsection{Habitable zone 1D climate model}\label{sec:expanded_climate_model}

We use the Virtual Planetary Laboratory (VPL) 1D radiative-convective climate
model\cite{meadows_habitability_2018,catling_atmospheric_2017} to generate
surface temperatures for an Earth-like planet at various pCO$_{2}$ and incident
fluxes. We consider incident fluxes between 1.05$S_{\oplus}$ and
$0.35S_{\oplus}$, the HZ limits for a Sun-like star
\cite{kopparapu_habitable_2013}, and atmospheric CO$_2$ partial pressures
between $10^{-6}$ and 10 bar. We assume the atmosphere is comprised of CO$_2$
and H$_2$O. If the CO$_2$ partial pressure is below 1 bar, we set the initial
atmospheric pressure to 1 bar and add N$_2$ to the atmosphere such that the
total surface pressure is 1 bar. We fix the stratospheric water vapor
concentration to the modern Earth value and follow the Manabe-Wetherald
relative humidity distribution in the troposphere with empirical constraints
based on the modern Earth \cite{manabe_thermal_1967}. 

\begin{figure}[htb!]
\centering
\includegraphics[width=10cm]{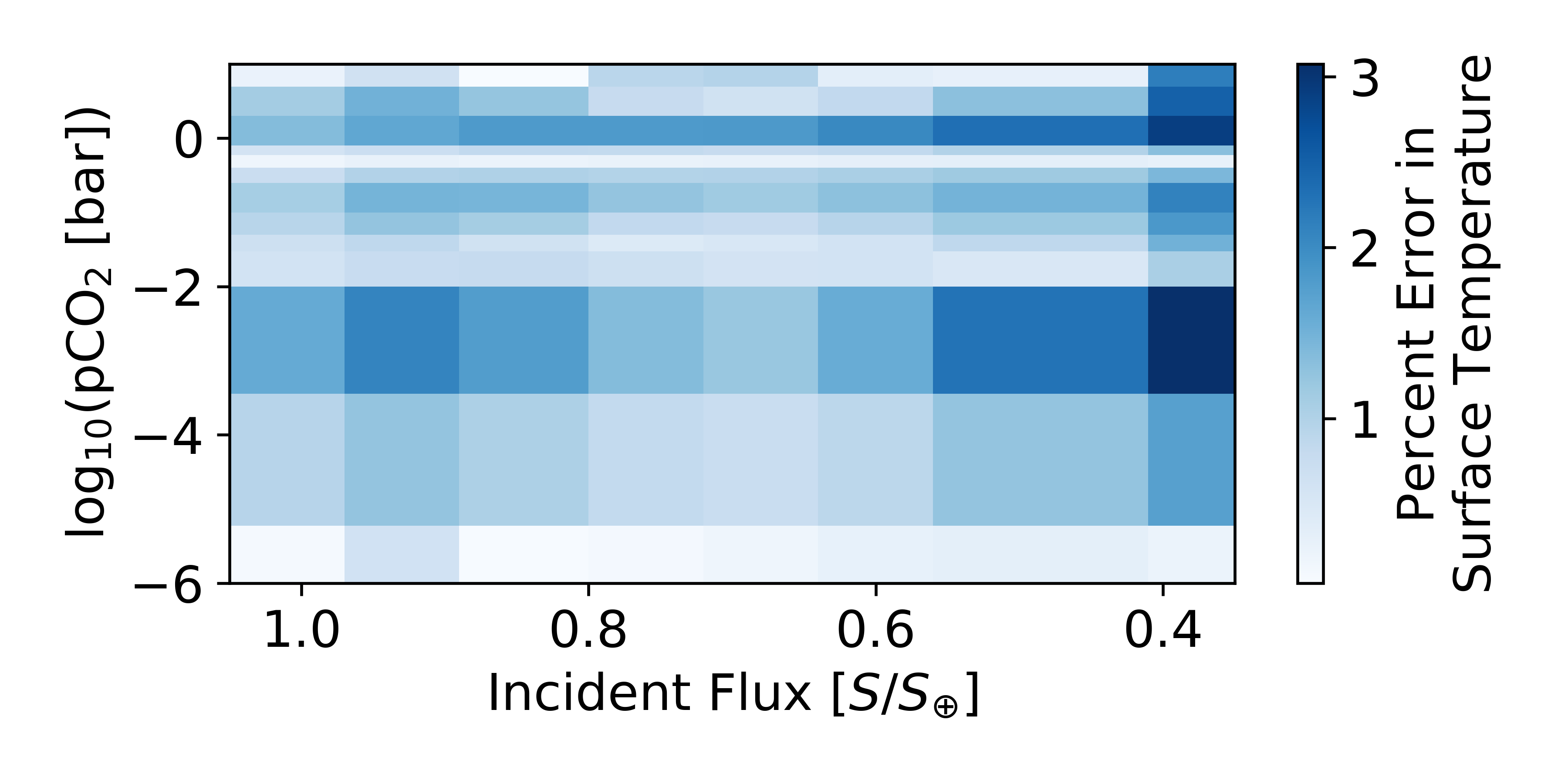}
\caption{The relative error between our 4$^{\textrm{th}}$-order polynomial fit and the
full 1D radiative-convective climate model. Our polynomial fit is valid between
1.05$S_{\oplus}$ and 0.35$S_{\oplus}$, where $S_{\oplus}$ is the Solar
constant. The polynomial fit is valid for atmospheric CO$_2$ abundances between
10$^{-6}$ and 10 bar. The surface temperatures predicted by the polynomial fit
reproduce the results of the 1D climate model. The maximum error in predicted
surface temperature between the polynomial fit and the 1D climate model is
${\sim}$3\%.}
\label{fig:clim_compared}
\end{figure}

We fit the surface temperature output, $T_{\textrm{s}}$ in K, from the climate model
with a 4$^{\textrm{th}}$ order polynomial in $\ln(\textrm{pCO}_{2})$ and normalized
stellar flux, as follows:
\begin{equation}\label{eqn:climate_fit_eqn}
\begin{split}
    T_{\textrm{s}}\left(S,\textrm{pCO}_{2}\right) =  4.809-222.0X-68.44X^{2}-6.737X^{3}-0.206X^{4} \\
    +1414XY+446.4X^{2}Y+44.41X^{3}Y+1.364X^{4}Y \\
    -2964XY^{2}-978.4X^{2}Y^{2}-98.86X^{3}Y^{2}-3.059X^{4}Y^{2} \\
    +2655XY^{3}+907.5X^{2}Y^{3}+92.87X^{3}Y^{3}+2.892X^{4}Y^{3} \\
    -868.4XY^{4}-304.6X^{2}Y^{4}-31.48X^{3}Y^{4}-0.985X^{4}Y^{4} \\
    +1045Y-1496Y^{2}+1064Y^{3}-281.1Y^{4}.
\end{split}
\end{equation}
Here, CO$_2$ partial pressure pCO$_2$ is in bar, $X=\ln(\textrm{pCO}_{2})$, and
$Y=S/S_{\oplus}$ is the incident flux, $S$, normalized to the solar constant,
$S_{\oplus}$. Figure \ref{fig:clim_compared} shows the agreement between the 1D
climate model and the polynomial fit used in this work.

\subsection{Numerical carbonate-silicate cycle
modeling}\label{sec:numerical_model}

To calculate the steady-state pCO$_2$ in the atmospheres of Earth-like planets
in the HZ, we use a weathering model that describes pCO$_2$ on the Earth
through
time\cite{krissansen-totton_constraining_2017,krissansen-totton_constraining_2018}.
We summarize the model below and highlight how the model in this work differs
from previous
implementations\cite{krissansen-totton_constraining_2017,krissansen-totton_constraining_2018}.
These previous implementations provide a comprehensive explanation and
justification of the model parameterizations, and empirical and theoretical
basis.  The model, as a Python script, is available in the Supplementary Data
and contains a complete description of the model equations and parameters (see
the file weathering\_model.py).

The weathering model balances the flux of outgassed CO$_2$ against the loss of
carbon due to continental and seafloor weathering, which result in
precipitation of carbonates in the ocean and seafloor pore space.
Quantitatively, for time $t$, this is described by time-dependent equations
where we normalize to the mass of the ocean, $M_{\textrm{o}}$ (nominally, an Earth
ocean, $1.35\times10^{21}$ kg): 

\begin{equation}\label{main_diff_eqs}
\begin{split}
      \frac{dC}{dt}&=\frac{F_{\textrm{out}}+F_{\textrm{carb}}-P_{\textrm{o}}-P_{\textrm{p}}}{M_{\textrm{o}}} \\ \frac{dA}{dt}&=2\cdot\frac{F_{\textrm{sil}}+F_{\textrm{carb}}+F_{\textrm{diss}}-P_{\textrm{o}}-P_{\textrm{p}}}{M_{\textrm{o}}}.   
\end{split}
\end{equation}
Here, $C$ is the non-organic carbon content of the atmosphere-ocean system in
mol C kg$^{-1}$, and $A$ is the carbonate alkalinity in mol equivalents (mol
eq). Carbonate alkalinity (henceforth alkalinity) is the charge-weighted
sum of the mol liter$^{-1}$ concentration of bicarbonate and carbonate anions,
[HCO$_3^{-}$] + 2[CO$_3^{2-}$]. $F_{\textrm{out}}$ is the global CO$_2$ outgassing flux,
$F_{\textrm{carb}}$ and $F_{\textrm{sil}}$ are the continental carbonate and silicate weathering
fluxes, $F_{\textrm{diss}}$ is the rate of seafloor basalt dissolution, and $P_{\textrm{p}}$ and
$P_{\textrm{o}}$ are the pore and ocean precipitation fluxes. The fluxes on the
right-hand side of equation \ref{main_diff_eqs} ($F_{\textrm{out}}$, $F_{\textrm{carb}}$,
$P_{\textrm{o}}$, $P_{\textrm{p}}$, $F_{\textrm{sil}}$, $F_{\textrm{carb}}$, and $F_{\textrm{diss}}$) are given in mol C
yr$^{-1}$ for $dC/dt$ and in mol eq yr$^{-1}$ for $dA/dt$.

The alkalinity that enters the ocean from weathering will balance a $+2$ charge
cation (e.g. Ca$^{++}$), which is why a factor of $2$ enters in the definition
of $dA/dt$ in equation \ref{main_diff_eqs}. Hence, geochemists often think of
alkalinity in terms of the balance of cations produced in weathering, principally
Ca$^{++}$. This reasoning arises because the weighted sum of carbonate and
bicarbonate concentrations must balance the charge of conservative cations
minus conservative anions (i.e., 2[Ca$^{++}$] + 2[Mg$^{++}$] + Na$^{+}$ +... -
[Cl$^{-}$] - ...), ignoring minor contributions from weak acid anions and water
dissociation. Weathering releases cations and carbon speciation adjusts to
ensure charge balance, so that the cation release is effectively equivalent to
carbonate alkalinity. 

To improve the rate of model convergence and range of model inputs over which
equation \ref{main_diff_eqs} converges, we do not consider the seafloor pore
space and atmosphere-ocean as separate systems. This differs from
previous versions of the
model\cite{krissansen-totton_constraining_2018}, which considered the
atmosphere-ocean and pore space independently. Rather, we approximate the
atmosphere-ocean and pore space as a single entity in equation
\ref{main_diff_eqs}. This simplification does not appreciably change the model
output for atmospheric CO$_2$ because we run the model to steady-state in all
cases, where the atmosphere-ocean and pore space reach approximate equilibrium.
In the next section, we present additional details on our model implementation
and discuss the agreement between our no-pore model and the original, two-box
model\cite{krissansen-totton_constraining_2018}.

A second modification is the range of incident stellar fluxes over which the
model can be run. Previously, the model described here was
used to study the Earth through time\cite{krissansen-totton_constraining_2018}
and thus only considered solar fluxes between $S_{\oplus}$ (the modern solar
constant) and early Earth's $0.7S_{\oplus}$ ($S_{\oplus}=1360$ W m$^{-2}$). We
extend that range to include the entire conservative HZ of a Sun-like star,
roughly 1.05$S_{\oplus}$ to 0.35$S_{\oplus}$ \cite{kopparapu_habitable_2013}.
We use equation \ref{eqn:climate_fit_eqn}, the 4$^{\textrm{th}}$-order polynomial fit to
a 1D climate model, to calculate surface temperatures throughout the HZ.  The
Bond albedo of the planet is calculated dynamically by the climate model and
thus included implicitly in our polynomial fit.

With the coupled climate and weathering model, we generate steady-state,
Earth-like planets by randomly sampling plausible initial model inputs. The
ranges for each parameter we consider are
representative of the Earth through
time\cite{krissansen-totton_constraining_2018} and shown in Table
\ref{tab:model_params}. These ranges represent very broad uncertainties of the
carbonate-silicate cycle on the Earth through time and so are appropriate for
Earth-like planets. We conservatively assume a uniform distribution for each
parameter range shown in Table \ref{tab:model_params}.

We parameterize the internal heat of an Earth-like planet conservatively using
the planet's age, ranging 0 to 10 Gyr, which is the approximate habitable
lifetime of an Earth-like planet around a Sun-like star
\cite{rushby_habitable_2013}. The equation for planetary heat relative to the
modern Earth, $Q$, is given by
\begin{equation}\label{internal_heat_eqn}
    Q=\left( 1 - \frac{4.5-\tau}{4.5} \right)^{-n_{\textrm{out}}}
\end{equation}
where $\tau$ is the age of the planet in Gyr, and $n_{\textrm{out}}$ is the scaling
exponent for internal heat, with a range given in Table 1.

The parameter ranges shown in Table \ref{tab:model_params}
represent the uncertainty of the carbonate-silicate weathering cycle on the
Earth through time \cite{krissansen-totton_constraining_2018}. Implicit in our
assumed parameter ranges is that continental land fraction, $f_{\textrm{land}}$, and
biological weathering fraction, $f_{\textrm{bio}}$, have increased from 0 when
the Earth formed to 1 on the modern Earth. Similarly, the relative internal
heat, $Q$, is assumed to be large when the Earth is young and unity for the
modern Earth.  Therefore, on the modern Earth, where $f_{\textrm{land}}=1$, $f_{\textrm{bio}}=1$,
and $Q=1$, the weathering rate is maximized and outgassing rate is relatively
small (see Methods, subsection
\nameref{sec:parameter_sensitivity} for a discussion on the importance of
these three parameters in our model). This is seen in Figure
\ref{fig:sim_exo_data}, where the modern Earth appears near the lower bound
for predicted pCO$_2$ in the HZ. If the continents on an exoplanet were
more easily weathered or outgassing much lower than on the modern Earth,
such exoplanets could have pCO$_2$ values well below the modern Earth value
shown in Figure \ref{fig:sim_exo_data}. We do not consider such exoplanets
in this model, so the results presented here are only applicable to planets
similar to the Earth through time.

Our model assumes that each simulated planet is habitable, i.e., it has a
stable, liquid surface ocean, a necessity for the carbonate-silicate cycle to
operate. For a mean surface temperature below 248 K, Earth-like planets are
likely completely frozen \cite{charnay_exploring_2013}, which we use as a lower
temperature bound in the model. While 248 K is below the freezing point of
water, it is a global mean surface temperature and 3D models show that the
range 248-273 K for this parameter does not preclude the existence of a liquid
ocean belt near the equator. At the other temperature extreme, a hot,
Earth-like planet can rapidly lose its surface oceans due to high atmospheric
water vapor concentrations that are photolyzed and subsequently lost to space.
This upper temperature bound on habitability occurs at ${\sim}355$ K
\cite{wolf_constraints_2017}. Above 355 K, Earth-like planets are unlikely to
remain habitable for more than ${\sim}1$ Gyr \cite{wolf_constraints_2017} and
cannot operate a carbonate-silicate cycle over geologic timescales. We use
these two temperature bounds, 248 K and 355 K, as the limits for habitability
in our model. Any modeled planet with a final surface temperature outside these
limits is uninhabitable and removed from our results. 

We limit HZ planets to those with pCO$_{2}$ below 10 bar. For most Earth-like
planets in the HZ, 10 bar of CO$_2$ results in planets with surface
temperatures well above 355 K, which are not habitable on long time scales. If
we impose a fixed stratospheric water vapor concentration in the 1D climate
model and modify the tropospheric water vapor concentration based on empirical
data from the modern Earth, we enable the 1D climate model to accurately model
habitable, Earth-like planets through much of the HZ. But in the outer HZ, with
more than ${\sim}$10 bar of CO$_2$, this assumption overestimates atmospheric
water vapor concentrations and leads to artificially warm planets, so we reject
such cases. Above ${\sim}$10 bar of CO$_2$ in the outer HZ, assuming a
saturated troposphere for water vapor, increasing atmospheric CO$_2$ may not
lead to additional warming\cite{kopparapu_habitable_2013}. Rather, the surface
cools in such scenarios because additional CO$_2$ leads to increased Rayleigh
scattering and no additional warming. Because Earth-like planets in the outer
HZ would be frozen and uninhabitable even with CO$_2$ partial pressures above
${\sim}$10 bar, we impose a 10 bar limit for CO$_2$ in the outer HZ. This limit
agrees with previous CO$_2$ limitations in coupled climate
and weathering models\cite{kadoya_conditions_2014}.

\subsection{Combined ocean and pore space model justification}\label{no_pore_justification}

The carbon cycle model used in this work was previously derived as a
 two-box model\cite{krissansen-totton_constraining_2018}, where the
atmosphere-ocean and the seafloor pore space were separated. In this work, we
combine the ocean-atmosphere and the pore space into a single unit. This
modification can be implemented in the original model\cite{krissansen-totton_constraining_2018} by assuming that the pH of the
pore space is the same as the pH of the ocean, and assuming that the alkalinity
and carbon content of the ocean and pore space are the same. The dissolution
and precipitation fluxes can then be calculated without treating the
ocean-atmosphere and the pore space as different systems. This modification
allows the model to converge quicker over a wider range of parameter
combinations. 

To validate our combined model, we ran the modern Earth through both the
original, two-box model\cite{krissansen-totton_constraining_2018} and our
modified model at 10 different incident fluxes between $S_{\oplus}$ and
0.7$S_{\oplus}$. The average error in predicted CO$_2$ values between our model
and the two-box model was 2.8\%, with a minimum error of 2.3\%, and a maximum
error of 3.6\%. Given the large uncertainties in model inputs (Table
\ref{tab:model_params}), the few percent error introduced by our simplified
model is unimportant.

For each parameter combination in our simplified model, we
start with the modern Earth then impose a step change for each model
parameter. We then run the simulation for 10 Gyr or until the system reaches
steady-state. We consider the model to have reached
steady-state when extrapolation of the rate of change of pCO$_2$ for 1 Gyr
changes pCO$_2$ by less than 1\%. Typically, the model converges within a few
Myr to a few tens of Myr. Rarely (2 of the 1200 planets simulated in this
work), parameter combinations will not reach steady-state after 10 Gyr.
Simulations with combinations of exceptionally high outgassing rates and low
CO$_2$ weathering rates can enter a regime were atmospheric CO$_2$ builds
without bound, never converging. Such model results are beyond the range of
validity of our model.

\subsection{Validity of carbon cycle parameterizations to exoplanets}\label{sec:parameter_sensitivity}

The parameterization of weathering in our model has been empirically validated
for the modern Earth \cite{walker_negative_1981,
krissansen-totton_constraining_2017, krissansen-totton_constraining_2018}. The
exponential temperature-dependence of continental weathering is a reasonable
approximation that agrees with field and lab measurements
\cite{walker_negative_1981} and can reproduce the climate results of more
complex models \cite{krissansen-totton_constraining_2017,
krissansen-totton_constraining_2018}. Similarly, the power-law parameterization
for the pCO$_2$-dependence of continental weathering agrees with data from the
modern Earth \cite{krissansen-totton_constraining_2017} and can even be
approximately derived from equilibrium chemistry arguments for an Earth-like
exoplanet \cite{graham_thermodynamic_2020}. The bulk geochemistry of rocky
exoplanets may be similar to Earth's \cite{doyle_oxygen_2019}, so we expect our
weathering parameterization to reasonably approximate
Earth-like planets in the HZ. However, uncertainties in how the
carbonate-silicate weathering cycle regulates climate on Earth persist
\cite{krissansen-totton_constraining_2018}, so the predicted variations in
pCO$_2$ in our model may not capture the true variability of pCO$_2$ in the HZ.
Below, we show that our broad parameterization of the carbonate-silicate
weathering cycle may encompass the plausible range of pCO$_2$ for the Earth
through time, but improved understanding the carbonate-silicate weathering
cycle may be necessary to know if such variations are indeed representative of
the Earth through time and applicable to Earth-like planets generally.

The rate of weathering depends strongly on the intrinsic features of a planet,
such as the CO$_2$ outgassing rate and the properties of its continents.
Changes in continental uplift rate, lithology, and configuration are
parameterized in our model through the $f_{\textrm{land}}$ and $f_{\textrm{bio}}$ terms. The
parameters $f_{\textrm{land}}$ and $f_{\textrm{bio}}$ linearly scale the weathering flux and
could analogously be considered a continental weatherability scaling
factor. For the ranges of $f_{\textrm{land}}$ and $f_{\textrm{bio}}$ considered in our model (see
Table \ref{tab:model_params}), changes in the continental
weatherability alone can generate pCO$_2$ values spanning ${\sim}4$ orders of
magnitude. This broad parameterization likely encompasses pCO$_2$ perturbations
due to continental weatherability changes caused by large volcanic
eruptions or changes in continental configuration. Indeed, the largest,
constrained change in pCO$_2$ due to such events on Earth may be closer to
${\sim}1$ order of magnitude, coeval with the eruption of the Siberian Traps
\cite{johansson_interplay_2018}.

\begin{figure}[htb!]
\centering
\includegraphics[width=10cm]{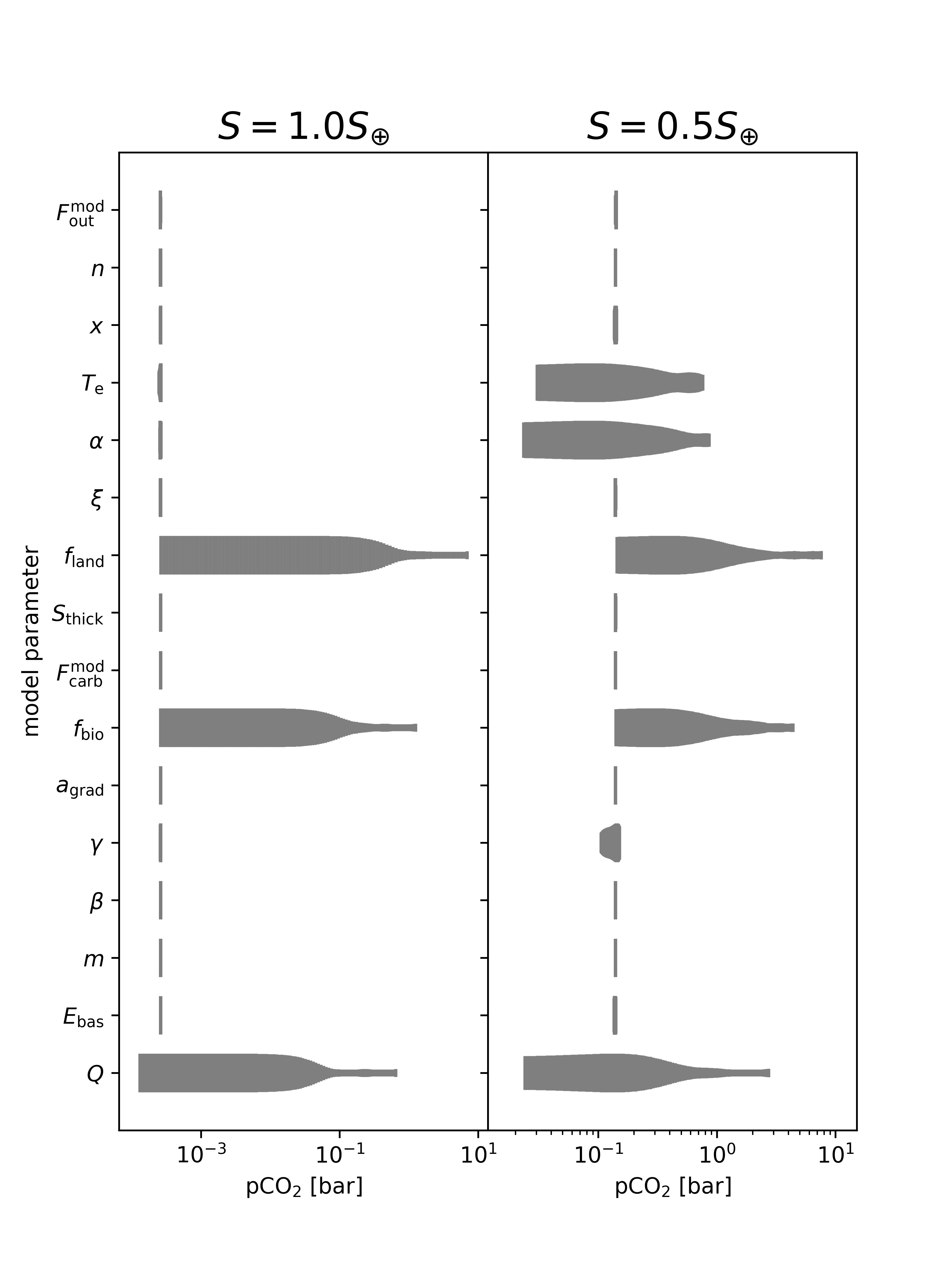}
\caption{The spread in steady-state pCO$_2$ from varying a single model
parameter. Each parameter in Table \ref{tab:model_params} is shown on the
vertical axis (note that $n_{\textrm{out}}$ and $\tau$ are incorporated
into $Q$, see equation \ref{internal_heat_eqn}). The left panel shows an
incident flux of $S=1.0 S_{\oplus}$. The right panel shows an incident flux
of $S=0.5 S_{\oplus}$. For each parameter, we held all other parameters
constant at the modern Earth value (see text) and randomly sampled 100
values for the parameter in question from uniform distributions of the
ranges given in Table \ref{tab:model_params}. The horizontal extent of the
gray shaded region shows the range of possible pCO$_2$ values when all
other parameters are fixed. The thickness of each gray shaded region shows
the relative abundance of steady-state planets at that pCO$_2$. The
thickest regions show maximum relative abundance, the thinnest regions show
the minimum relative abundance.  No surface temperature limits on
habitability were imposed for the simulated planets. At low pCO$_2$, three
parameters ($f_{\textrm{land}}$, $f_{\textrm{bio}}$, and $Q$) dominate the
spread in pCO$_2$. At higher pCO$_2$, the temperature- and
pCO$_2$-dependence of continental silicate weathering ($T_{\textrm{e}}$ and
$\alpha$) are also important. }
\label{fig:sm_params}
\end{figure}

The importance of continental weatherability ($f_{\textrm{land}}$ and
$f_{\textrm{bio}}$) on pCO$_2$, relative to other parameters, is shown in Figure
\ref{fig:sm_params}. Figure \ref{fig:sm_params} was generated by sampling
uniform distributions for each model parameter shown in Table
\ref{tab:model_params} across its listed range. When one parameter was varied,
all other parameters were held constant at their modern Earth value, which we
define as: $F_{\textrm{mod}}^{\textrm{out}}=6$ Tmol C yr$^{-1}$, $n=1.75$, $x=1$, $T_{\textrm{e}}=25$ K,
$\alpha=0.3$, $\xi = 0.3$, $f_{\textrm{land}}=1$, $S_{\textrm{thick}}=1$, $F^{\textrm{mod}}_{\textrm{carb}}=10$
Tmol C yr$^{-1}$, $f_{\textrm{bio}}=1$, $a_{\textrm{grad}}=1.075$, $\gamma = 0.2$, $\beta = 0.1$,
$m=1.5$, $E_{\textrm{bas}}=90$ kJ mol$^{-1}$, and $Q=1$. Note that we incorporate
$n_{\textrm{out}}$ and $\tau$ from Table \ref{tab:model_params} into $Q$, the internal
heat (see equation \ref{internal_heat_eqn}), which is the parameter of
interest. We show two different values for $S$ in Figure \ref{fig:sm_params},
$S=1.0 S_{\oplus}$ in the left panel and $S=0.5 S_{\oplus}$ in the right panel.
For both values of $S$, Figure \ref{fig:sm_params} shows that variations in
$f_{\textrm{land}}$ and $f_{\textrm{bio}}$ alone can alter pCO$_2$ by orders of magnitude.

The internal heat of the planet, $Q$, plays a similarly important role in
setting pCO$_2$. The rate of CO$_2$ outgassing is determined by $Q$ and our
broad parameterization of $Q$ allows pCO$_2$ to vary by orders of magnitude
throughout the HZ, as shown in Figure \ref{fig:sm_params}.

\begin{figure}[htb!]
\centering
\includegraphics[width=10cm]{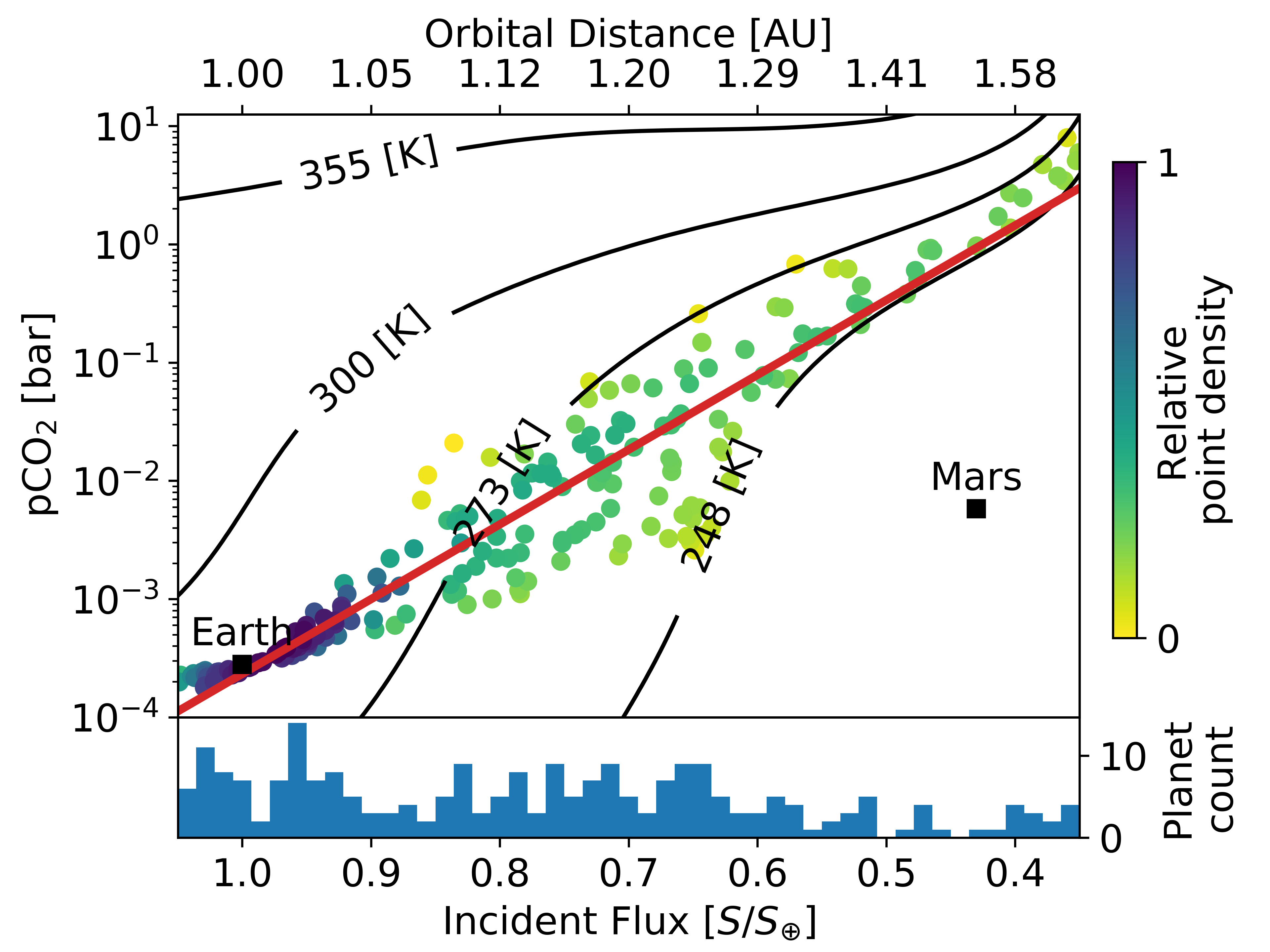}
\caption{The expected distribution of stable, steady-state pCO$_2$ on
Earth-like planets if $Q$, $f_{\textrm{land}}$, and $f_{\textrm{bio}}$ are
fixed to 1, i.e., modern Earth values. Except for fixing
$Q=f_{\textrm{land}}=f_{\textrm{bio}}=1$, this figure is generated
identically to Figure \ref{fig:sim_exo_data}. The spread in pCO$_2$ in the
outer HZ is due to the temperature- and pCO$_2$-dependence of continental
weathering ($T_{\textrm{e}}$ and $\alpha$). This is expected from equation
\ref{eqn:weathering_temp_dependence_full}, which shows that
$T_{\textrm{e}}$ and $\alpha$ will be increasingly influential as pCO$_2$
and surface temperature deviate from the modern Earth values, as discussed
in the Results, subsection \nameref{sec:theory}.  Without
changes in $Q$, $f_{\textrm{land}}$, and $f_{\textrm{bio}}$, there is little
spread in pCO$_2$ in the inner HZ.}
\label{fig:exo_data_fixed_params}
\end{figure}

The rate of CO$_2$ outgassing and continental weatherability drive the majority
of the spread in pCO$_2$ shown in Figure \ref{fig:sim_exo_data}. This is
readily seen in Figure \ref{fig:exo_data_fixed_params}, which shows the results
of 300 random parameter combinations from uniform distributions of the
parameters in Table \ref{tab:model_params} except for $Q$, $f_{\textrm{land}}$,
and $f_{\textrm{bio}}$, which were all fixed to 1. Of the 300 parameter
combinations, 235 remained above 248 K and are shown in Figure
\ref{fig:exo_data_fixed_params}.  Comparing Figure
\ref{fig:exo_data_fixed_params} to Figure \ref{fig:sim_exo_data}, it is readily
apparent that the broad uncertainty in pCO$_2$ from our results is due to
variations in intrinsic planetary properties ($Q$, $f_{\textrm{land}}$, and
$f_{\textrm{bio}}$) rather than uncertainties in the tuning parameters of our
carbon cycle parameterization. 

The outgassing rate and continental properties of habitable exoplanets remain
unknown. Thus, our broad parameterization of those terms, which align with
possible conditions on Earth throughout its history, are a reasonable
approximation. If an Earth-like, carbonate-silicate weathering cycle is common
on habitable planets, then these parameters may largely determine pCO$_2$ on
such planets and generate a range for pCO$_2$ at a given orbital distance
similar to that shown in Figure \ref{fig:sim_exo_data}.

\section*{Data Availability}
The data used in this work is available in the Supplementary Data. Our
model code depends on the location of the data directory, so the data and model
are provided together in a single, zipped file.

\section*{Code Availability}
The code used to generate the data and figures for this work is available in
the Supplementary Data.

\clearpage


\section*{Acknowledgements}
We would like to thank Nicholas Wogan for his constructive suggestions on our
initial manuscript. We also thank NASA's Virtual Planetary Laboratory (grant
80NSSC18K0829) at the University of Washington and the NASA Pathways Program
for funding this work. J.K.T. was supported by NASA through the NASA Hubble
Fellowship grant HF2-51437 awarded by the Space Telescope Science Institute,
which is operated by the Association of Universities for Research in Astronomy,
Inc., for NASA, under contract NAS5-26555.

\end{document}